\documentclass{iauc}
\usepackage{graphicx}
\pubyear{2005}
\volume{199}
\pagerange{1--}
\jname{Probing Galaxies through Quasar Absorption Lines}
\editors{P. R. Williams, C. Shu, and B. M\'{e}nard, eds.}
\title[Chemical Abundances in SFG and DLA] %% give here short title %% 
{Chemical Abundances in Star-forming Galaxies and Damped Lyman Alpha Systems}

\author[Schulte-Ladbeck, K\"onig \& Cherinka]   %% give here short author list %%
{Regina E. Schulte-Ladbeck, Brigitte K\"onig, Brian Cherinka
  }

\affiliation{Department of Physics and Astronomy, University of Pittsburgh,
Pittsburgh, PA 15260, USA}

\begin{document}

\maketitle

\begin{abstract}

We investigate the chemical abundances of local star-forming galaxies
which cause Damped Lyman Alpha lines. A metallicity versus redshift
diagram is constructed, on which the chemical abundances of
low-redshift star-forming galaxy populations are compared with those of
high-redshift Damped Lyman Alpha systems. We disucss two types of
experiments on individual star-forming galaxies. In the first, the Damped Lyman
Alpha line is created against an internal ultraviolet light source
generated by a star-forming cluster or a supernova explosion. In the
second, the Damped Lyman Alpha line is seen against a background
Quasar. The metallicities measured from ionized gas in the
star-forming regions, and neutral gas in the Damped Lyman Alpha systems, are
compared with one another on a case-by-case basis. We highlight the
occurrence of the star-forming galaxy/Quasar pair SBS~1543+593/HS~1543+5921,
where the emission- and absorption-line derived abundances give the
same result. We argue that we therefore can in principle, 
interpret Damped Lyman Alpha system metallicities as an extension of
star-forming galaxy metallicities to higher redshifts, supporting that
gas-rich galaxies had lower chemical abundances when the were younger.

\end{abstract}

\keywords{Galaxies: abundances, ISM, quasars: absorption lines}
%% add here a maximum of 10 keywords, to be taken form the file <Keywords.txt>.

\firstsection % if your document starts with a section,
\section{Introduction}
The chemical evolution of the Universe is due to two
channels. Big-bang nucleosynthesis generated several light chemical
elements during ``the first three minutes". In the following 13\,Gyr,
the metals have been synthesized by stars in galaxies; a process which
continues until the present epoch. The ratios of certain metals, for
example, those of the $\alpha$-to-iron-group elements, can be used to
constrain the evolution of galaxies.

Gas-phase chemical abundances in the high-redshift Universe have been
obtained by the study of QSO absorption lines, and traditionally refer
to iron-peak elements (e.g., \cite[Prochaska at
al. 2003]{prochaska2003}). It is generally thought that absorbers with
high neutral hydrogen column densities, such as Damped Lyman Alpha
(DLA) systems and sub-DLAs, originate in gas-rich galaxies. Typical
impact parameters needed to reach the DLA-defining column density, N,
of 2 $\times 10^{20}{\rm cm}^{-2}$, are expected to be of the order of
a few 10\,kpc based on the HI sizes of local gas-rich galaxies
(\cite[Rosenberg \& Schneider 2003]{rosenberg2003}). However, the
connection between DLAs and galaxies is far from clear. For instance,
the very low metallicity of high-redshift DLAs has been interpreted to
indicate that DLA galaxies are preferentially dwarf galaxies, in
which case they would not be representative of the local field galaxy
population (e.g., \cite[Kulkarni et al. 2005]{kulkarni2005}).

\begin{figure}[h]
\vspace{0.5cm}
\parbox{0.6\textwidth}{\includegraphics[width=0.6\textwidth]{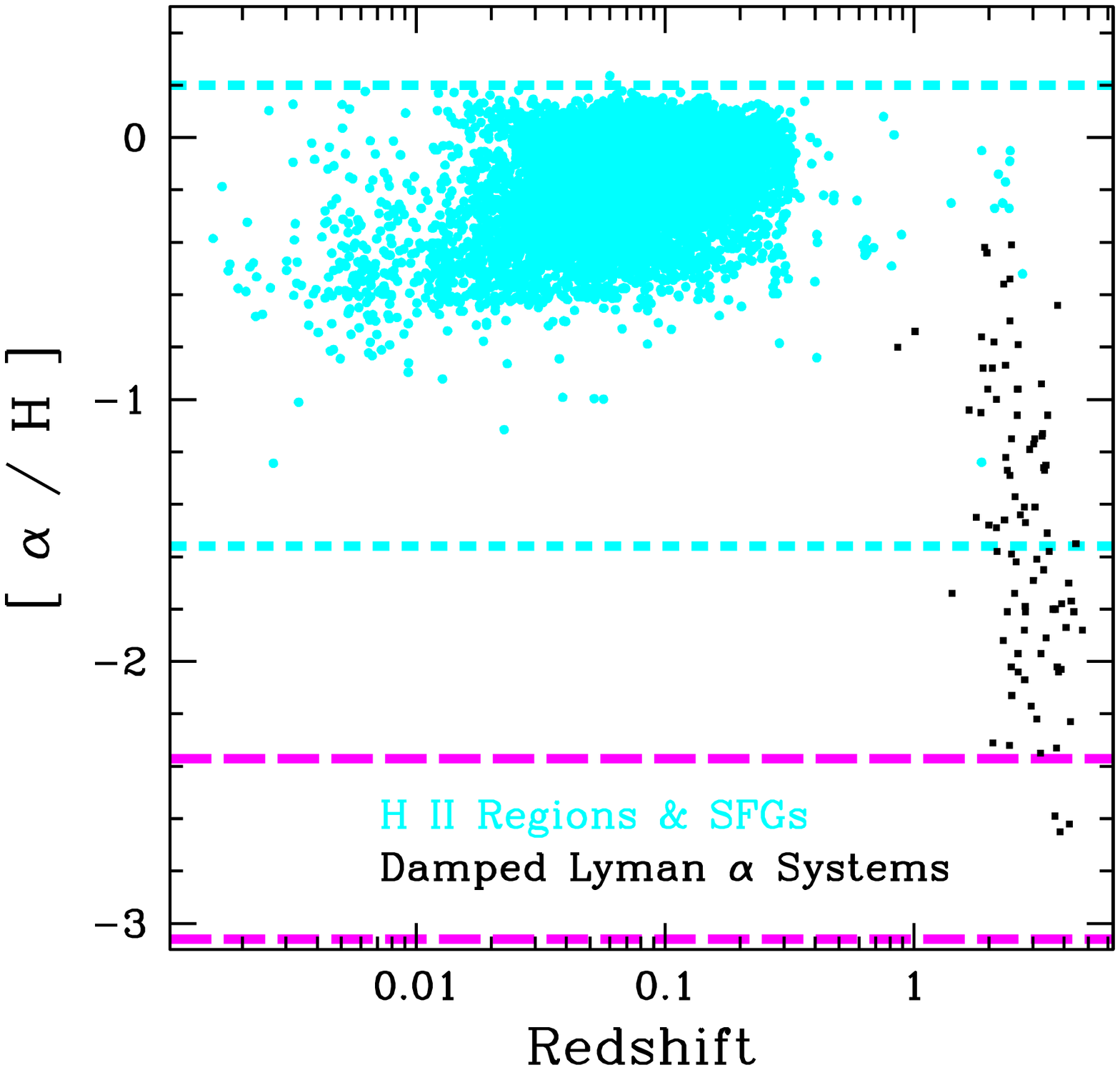}}
\hspace{0.5cm}
\vspace{-1cm}
\parbox{0.4\textwidth}{\caption{Metallicity as a function of
redshift. The bright circles are values of ${\rm [O/H]}_{\rm II}$\ for
HII regions and SFGs from the SDSS. These are extended to redshifts
above 0.3 by using data from \cite{shapley2004}, \cite{steidel2004},
\cite{koo1994}, \cite{kobulnicky1999}, \cite{lb2003},
\cite{liang2004}, \cite{rigo2000}, \cite{maier2004},
\cite{contini2002}, \cite{cardiel2003}, and \cite{teplitz2000}. The
black squares are values of ${\rm [O,S,Si/H]}_{\rm I}$\ from the
compilation of \cite{prochaska2003d}. The short-dashed lines show the
maximum and minimum values of oxygen abundances in local HII regions
and SFGs based on the direct, or T$_e$ method (\cite[Izotov \& Thuan
1999]{IT1999}, \cite[Castellanos, D\'iaz \& Terlevich
2002]{Cast2002}). The long-dashed lines are the theoretical
expectations of oxygen enrichment from the first supernovae
(\cite[Salvaterra \& Ferrara 2003]{salvaterra2003}), assuming a
non-rotating (lower line) and a rotating (upper line) Population~III
progenitor.}}
\label{fig1}
\end{figure}

Only at low redshifts can the galaxies responsible for causing DLAs be
observed directly. We discuss here a few cases for which detailed
studies are in progress. Chemical properties of galaxies at low
redshifts are based on measurements of emission lines from
photo-ionized gas, and refer to the $\alpha$-capture element oxygen
(e.g., \cite[Garnett 2004]{garnett2004}). Figure~\ref{fig1}
illustrates the local $\alpha$-element abundances derived from
emission-lines, and the high-redshift abundances derived for DLAs from
absorption lines, showing only those for which $\alpha$- rather than
Fe-peak element abundances are available. We also indicate the maximum
and minimum oxygen abundances seen in the local universe, and note
that a) there are no local galaxies known with abundances as low as
those of the lowest-metallicity DLAs; and b) there are no DLAs known
which reach super-solar abundances known locally to occur in the most
luminous SFGs. The chemical abundances of Lyman Break Galaxies (LBGs)
and of DLAs are beginning to show some overlap at redshift of about 2,
where emission-line abundances of LBGs have recently become available
(e.g. \cite[Teplitz et al. 2000]{teplitz2000}, \cite[Shapley et
al. 2004]{shapley2004}). LBGs are luminous star-forming galaxies
(SFG), which are presumed to be offset from the local
luminosity-metallicity relation because of their youth. \footnote{We
  use a cosmology with H$_0 = 70\,{\rm km}\,{\rm s}^{-1}{\rm
    Mpc}^{-1}$, $\Omega_m=0.3$, and $\Omega_\Lambda=0.7$. All
  literature data were converted to this cosmology when needed.}

In order to compare the properties of SFGs with those of high-redshift
DLAs, we need to answer the question whether emission- and
absorption-line techniques yield concordant values for chemical
abundances. The experiment requires the use of a SFG, a background
light source, and a high neutral H column density on its path. For
some time, the neutral gas in SFGs has been probed against internal
star clusters. More recently, the studies have been expanded to bona
fide DLA galaxies using SFG-QSO pairs.

\section{SFGs causing DLAs in their own spectra}
\begin{figure}[h]
\vspace{0.5cm} 
\hfill
  \begin{minipage}[t]{.49\textwidth}
    \begin{center}  
\includegraphics[width=\textwidth]{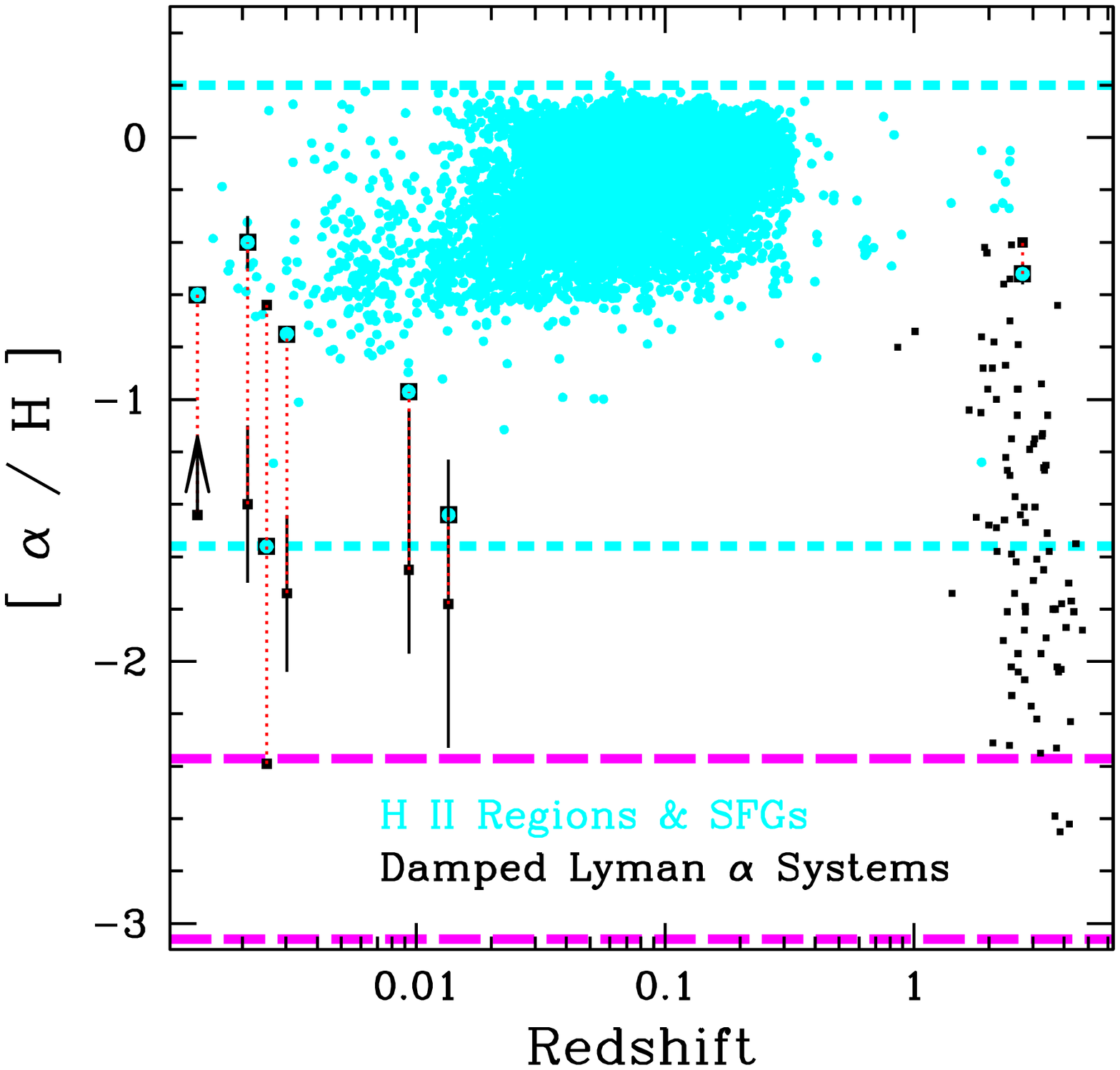}
%\caption{Metallicity as a function of redshift as in
%Fig.~1. Overplotted are the abundances of SFGs from Table~1, which
%display DLAs in their own spectra.}

    \end{center}
  \end{minipage}
  \hfill
  \begin{minipage}[t]{.49\textwidth}
    \begin{center}  
\includegraphics[width=\textwidth]{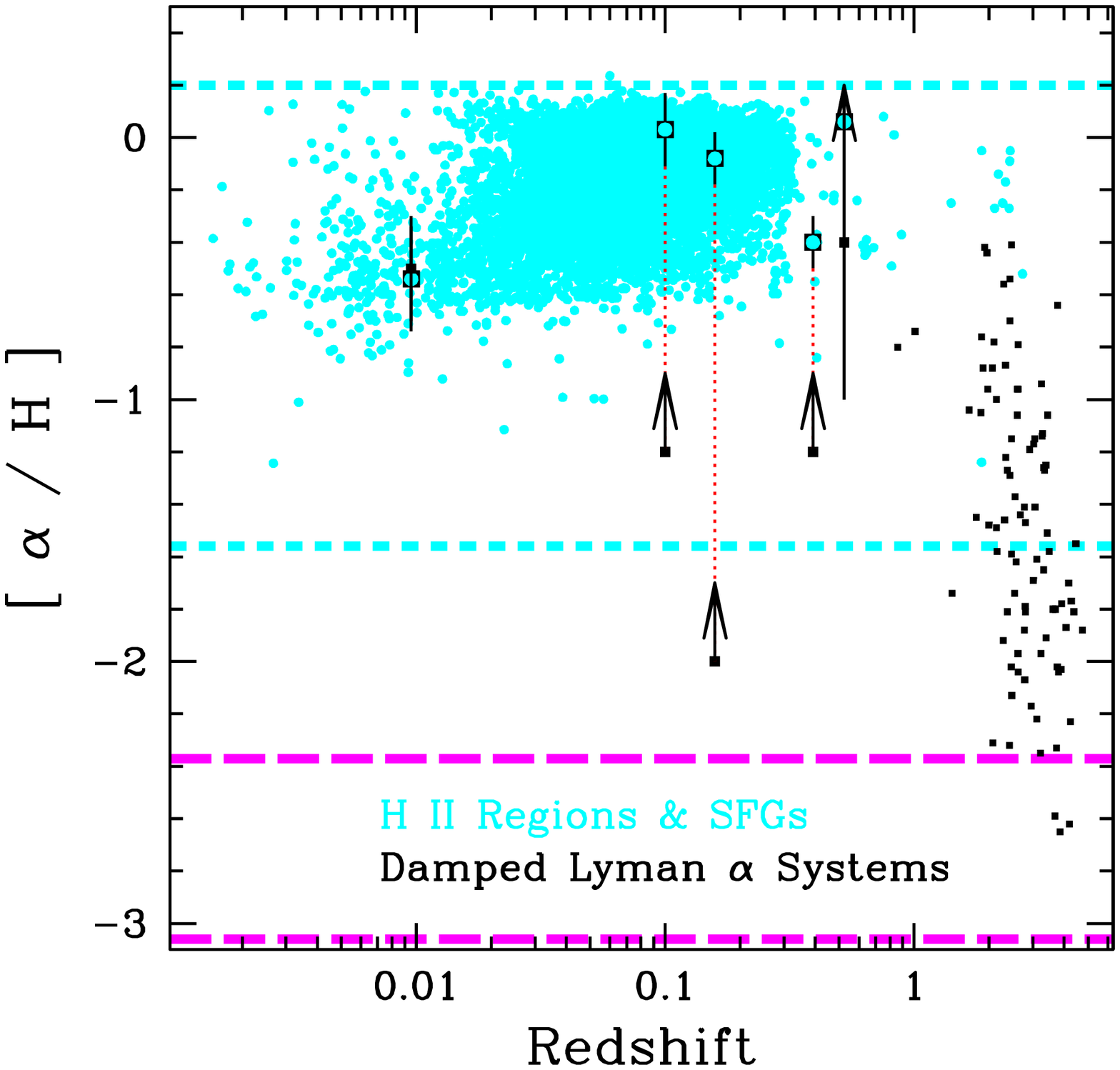}
%\caption{Metallicity as a function of redshift as in
%Fig.~1. Overplotted are the abundances of SFGs from Table~2, as
%well as the unambiguous abundances (or abundance limits) determined on the
%sightline to the QSO.}

    \end{center}
  \end{minipage}
\hfill
\vspace{-2cm}
\caption{Left -- Metallicity as a function of redshift as in
Fig.~1. Overplotted are the abundances of SFGs from Tab.~1, which
display DLAs in their own spectra.
For each SFG which is also a DLA, the bright symbols mark the ionized gas
abundance; and they are connected by dotted lines, to dark squares which
indicate the neutral gas abundance. For I~Zw~18, the extrema of neutral gas
metallicities allowed by the analyses of two teams, are shown.
\newline 
{\bf Figure~3.} Right -- Overplotted here are the abundances of SFGs
from Tab.~2, and the corresponding unambiguous abundances (or
abundance limits) determined on the sightline to the QSO.}  \hfill
\end{figure}

Table~\ref{tab1} provides a list of local SFGs exhibiting DLA lines in
their own, UV spectra. Chemical abundances are referenced to the solar
values of \cite{holweger2001}. For oxygen this is $12+\log {\rm O/H} =
8.736\pm0.078$.  The neutral gas-phase abundances are for oxygen, with
the exception of MS~1512-cB58, where they refer to the average of
several $\alpha$ elements. Oxygen abundances in the ionized gas phase
are given as well. We illustrate how the abundances of these SFGs
compare to other, low-redshift SFGs, and to high-redshift DLAs, on
Fig.~2.

\begin{table}
\caption{SFGs causing Lyman alpha lines in their own spectra} 
\label{tab1}
\begin{tabular}{lcccccccc}
\hline
Galaxy  & z  & $\log({\rm N_{HI}})$ & Type & M$_B$ & L & SFR & [O/H]$_{\rm II}$ &
[O/H]$_{\rm I}$\\
  &   &   &  &  & [$L^{\ast}$] & [M$_\odot$yr$^{-1}$] &  & \\
\hline
NGC 625      & 0.001321 & 20.5  & DIRR      & $-16.3$  & 0.03     & 0.05    & $-0.6^{12} $ & $>-1.4^{11}$ \\   
NGC 1705     & 0.00210  & 20.20 & BCD       & $-15.8$  & 0.02     & 0.03    & $-0.4^5 $ & $-1.4^4$ \\
I Zw 18      & 0.00250  & 21.34 & BCD       & $-14.3$  & 0.004    & 0.09    & $-1.56^1$ & $-1.4^6$, $-2.1^7$ \\
Mrk 59       & 0.003631 & 20.8  & BCD       & $-17.5$  & 0.08     & 1.5     & $-0.75^1$ & $-1.7^8$ \\
I Zw 36      & 0.0094   & 21.30 & BCD       & $<-13.9$ & $>$0.003 & $>$0.03 & $-0.97^1$ & $-1.7^9$ \\
SBS 0335-052 & 0.013486 & 21.85 & BCD       & $-17.1$  & 0.04     & 1.3     & $-1.44^1$ & $-1.8^{10}$  \\
MS 1512-cB58 & 2.7276   & 20.85 & LBG       & $-22.0$  & 5.3      & $>100$  & $-0.52^2$ & $-0.4^{3}$ \\
\hline
\multicolumn{9}{l}{$^1$ \cite{IT1999}, $^2$ \cite{teplitz2000}, $^3$ \cite{pettini2002}
  , $^4$ \cite{heckman2001},}\\
\multicolumn{9}{l}{$^5$ \cite{sb1994}, $^6$ \cite{aloisi2003}, $^7$
  \cite{lde2004},}\\
\multicolumn{9}{l}{$^8$ \cite{IT2002}, $^9$ \cite{lb2004}, $^{10}$
  \cite{thuan2005},}\\
\multicolumn{9}{l}{$^{11}$ \cite{cannon2005}, $^{12}$ \cite{skillman2003}
}

\end{tabular}
\end{table}

The advantage of examining DLAs produced within SFGs by their own
starburst clusters, is that the neutral and the ionized gas are probed
on the exact same sightline. A disadvantage of this technique, is that
the background sources are extended. The sightline also, traverses
some ionized gas.

So far, six out of seven SFGs studied with this technique are dwarf
galaxies with small star-formation rates (SFR). In general, in these
galaxies, ${\rm [O/H]_{II}}$ -- ${\rm [O/H]_I} >0$ (although the case of
I~Zw~18 is controversial). The seventh galaxy studied is the luminous,
gravitationally-lensed, high-redshift LBG MS~1512-cB58. Here,
[O/H]$_{\rm II}$ and [O/H]$_{\rm I}$ are in close agreement.

The discovery of DLA lines in the spectra of gamma ray burst (GRB)
source afterglows has recently opened up a new window on neutral gas
metallicities at high redshift (e.g., \cite[Vreeswijk et al.,
2004]{vreeswijk2004}). GRBs are thought to originate from supernova
explosions in star-forming regions; and indeed, some GRBs exhibit
Lyman $\alpha$ emission lines in their DLA troughs. At this time,
abundance analyses of the ionized gas in GRB-DLAs are not available;
thus none of these objects are listed in Table~1.

\section{SFGs causing DLAs in the spectra of background QSOs}

\cite{sl2005} recently studied the $\alpha$-element
abundances of the DLA galaxy SBS~1543+593 in emission and
absorption. The background QSO HS~1543+5921 has an impact parameter of
2.4'', and intercepts the dwarf galaxy's disk close to its
center. SBS~1543+593 offers an excellent opportunity to directly
compare element abundances inferred from cool interstellar gas (DLAs)
and ionized gas (SFGs). None of the previously imaged DLA galaxies
resolves to show individual HII regions. In none of them, does the
sightline to the QSO intercept the disk of the galaxy close to its
center, eliminating concerns over disk metallicity gradients.

The brightest HII region in SBS~1543+593 yields ${\rm [O/H]}_{\rm II}
= -0.54\pm0.20$\ and ${\rm [S/H]}_{\rm II} = -0.27\pm0.30$ (where the
solar S abundance ($7.20\pm0.06$) is referenced to \cite[Grevesse \&
Sauval 1998]{grevesse1998}). HST/FOS data reveal ${\rm [O/H]}_{\rm
I}>-2.14$, and HST/STIS data give ${\rm [S/H]_I} =
-0.50\pm0.33$. Within the errors these four values are the same. Also,
log ${\rm (N/O)}_{\rm II} = -1.40^{+0.20}_{-0.30}$ is within the range
of $-2.0 \lesssim \log {\rm(N/S)_I} \lesssim -0.8$, suggesting
agreement.

%\begin{figure}[h]
%\includegraphics[width=\textwidth]{rsl3.ps}
%\caption{}
%\label{fig3}
%\end{figure}
On Fig.~3 we show ${\rm [S/H]_I}$ for the neutral-gas abundance, and
${\rm [O/H]}_{\rm II}$ for the ionized-gas abundance. Here is one DLA
for which we can demonstrate that in principle, emission- and
absorption-line techniques give the same results when chemical
elements with similar nucleosynthetic origins are compared at similar
locations within a DLA galaxy. This result validates in principle, the
comparisons between the two types of objects, SFGs and DLAs, on the
metallicity versus redshift diagram. SBS~1543+593 thus supports the
case that metal enrichment has taken place in the gas-rich,
star-forming galaxy population between redshifts of 5 and 0. In
practice, galaxy metallicity gradients will be important for a
detailed comparison between SFG and DLA metallicities (see
e.g. \cite[Christensen et al. 2005]{christensen2005}, \cite[Chen et
al. 2005]{chen2005}).

\begin{table}
\caption{SFGs causing Lyman alpha lines in background QSO spectra} 
\label{tab2}
{\tiny
\begin{tabular}{lcrccccccccl}
\hline 
QSO & b & Galaxy & z & $\log(N_{\rm HI})$ & Type & M$_B$ & L & SFR & [O/H]$_{\rm II}$ & [O/H]$_{\rm I}$ & $\alpha$, other \\
 & [kpc] & & & & & &[$L^{\ast}$] & [M$_\odot$yr$^{-1}$] & & & than O\\ 
\hline 
HS~1543+5921 & 0.5    & SBS~1543+593     & 0.0096 & 20.3       & Sm  & $-16.8$ & 0.04 & 0.006            & $-0.59^a$    & $-0.5^*$ & S, Si\\ 
Q1543+489    & 107    & Galaxy 5         & 0.0382 & 18.4       & S   & $-17.9$ & 0.12 & 0.37             & $-0.06^f$    & $>-0.8$? & S, Mg \\
PHL 909      & 140    & SDSS J005719$^+$ & 0.0822 & $\leq$17.7 & Sc  & $-19.7$ & 0.6  & 1.5              & $-0.22^g$    & ...     & \\ 
%OI 363      & $<$3.4 & jet \& arm       & 0.0912 & $21.00\pm0.15$ &     &         &      & 0.01             &              & $>-2.9$ & S, Mg, Si\\
PKS 0439-433 & 7      & Object 1         & 0.1012 & 19.7       & Sab & $-20.2$ & 1.0  & 0.4              & $ 0.03^b$    & $>-1.2$ & Mg, Si\\ 
PHL 909      & 19     & SDSS J005710     & 0.1237 & 18.58      & Sa  & $-18.5$ & 0.2  & 1.1              & $-0.15^h$    & $>-1.2$? & \\ 
PHL 1226     & 17.6   & G4               & 0.1602 & 19.7       & S   & $-20.0$ & 0.8  & 0.5              & $-0.08^d$    & $>-2.0$ & \\ 
Q1209+107    & 38     & Galaxy 19/8      & 0.3930 & 19.54      & Int & $-18.5$ & 0.2  & 1.7              & $-0.4^{c,e}$ & $>-1.2$ & \\ 
%PKS 2128-12 & 46     & Galaxy 1         & 0.4299 & $18.89\pm0.25$ & S   &         &      & 2.86             &              & $>-1.1$ & S\\ 
%AO 0235+164 & 7--40  & \#1, 2, 3        & 0.524  & $21.72\pm0.12$ &     &         &      & 1.32, 4.86, 1.21 & $-0.24^b$    & $>-3.2$ & S, Mg, Si\\ 
B2 0827+243  & 36     & G1               & 0.525  & 20.3       & S   & $-19.8$ & 0.69 & 0.08             & $>$0.06$^b$  & $>-2.8$ & S, Si\\ 
\hline
\multicolumn{11}{l}{$^a$\cite{sl2005}, $^b$ \cite{chen2005}, $^c$
\cite{cristiani1987}, $^d$\cite{christensen2005}}\\
\multicolumn{11}{l}{$^e$ \cite{yanny1990}, $^f$ \cite{bowen1997}, $^g$
  this work}\\
\multicolumn{11}{l}{The [O/H]$_{\rm I}$ values were derived assuming
the optically thin case and applying the approximation formula from
}\\
\multicolumn{11}{l}{\cite{petitjean1998}. $^*$ [S/H], $^+$ also SDSS~J154519} 
%\enddata
%\tablenotetext{a}{Schulte-Ladbeck et al. (2004), }
%\tablenotetext{b}{Chen, Kennicutt \& Rauch (2004), }
%\tablenotetext{c}{Cristiani (1987)}
%\tablenotetext{d}{Christensen et al. (2005)}
%\tablenotetext{e}{Janny (1990)}
%\tablecomments{$[O/H]_{\rm I}$ lower limits are determined measuring
%  the EW and assuming the optically thin case. But it is known that this
%  line is often saturated.} 
\end{tabular}
}
\end{table}

Table~\ref{tab2} summarizes the result for a few additional SFG/QSO
pairs, for which we were able to list $\alpha$-element abundances in
the ionized phase, and derive abundance limits in the neutral
phase. We find that in two sub-DLAs (G4/PHL~1226 and
Object1/PKS~0439-433), the QSO intercepts the disk just at its outer
radius. Here [$\alpha$/H]$_{\rm I}$ is consistent with expectations of
the local disk metallicity gradient of \cite{ferguson1998}. As the
impact parameter grows, this extrapolation must eventually become
invalid. We note that the present sample traces $\alpha$ elements to a
distance of about 100\,kpc from SFGs.

\cite{chen2005} have argued that low-redshift DLAs can be explained
via a combination of gas cross section selection and metallicity
gradients. Galaxy metallicity gradients have only been investigated
locally. We do not have any insight into how they may evolve with
redshift. One point in particular, is worth pondering. It is now
becoming evident that the optical sizes of galaxies grow smaller with
increasing redshift (\cite[Bouwens et al. 2004]{bouwens2004}). This
might suggest an as yet little explored bias in the observed
metallicity-redshift dependence of DLAs.

\section{Conclusions}

$\bullet$~ Emission and absorption diagnostics of one SFG/QSO pair
giving rise to a DLA, are shown to give the same $\alpha$-element
abundances for the ionized gas in the SFG and the neutral gas in the
DLA.

$\bullet$~ Emission- and absorption-derived abundance offsets in
SFG/QSO pairs in which the impact parameter is less than or equal to
the optical radius of the galaxy, so far appear consistent with the
expectation of a disk metallicity gradient.

$\bullet$~ High-redshift SFGs and DLAs exhibit lower abundances than
their low-redshift counterparts, indicating that gas-rich galaxies had
lower abundances when they were younger.

\begin{acknowledgments}

The paper makes use of publicly available SDSS data. The SDSS website
is http://www.sdss.org/. Chris Miller is thanked for providing the
catalog of line fluxes from which the SDSS HII-region and
SFG metallicities were derived. This paper was funded in part by NASA
HST archival project 10282. We thank the School of Arts \& Sciences
for support.

\end{acknowledgments}

\end{document}